\documentclass[a4paper,12pt]{article}
\usepackage[a4paper, margin=1in]{geometry}
\usepackage{amsmath}
\usepackage{amssymb}
\usepackage{mdframed}
\usepackage{titlesec}
\usepackage{blindtext}
\usepackage{tikz}
\usepackage{booktabs}
\usepackage{adjustbox}
\usetikzlibrary{decorations.pathreplacing}

\begin{document}
\title{Enhancing TreePIR for a Single-Server Setting via Resampling}
\author{Elian Morel}
\date{}
\maketitle
\section{Introduction}
In \textbf{Private Information Retrieval} (PIR), a server or multiple servers hold a public database \textbf{DB} with \( n \) entries, and a client wishes to retrieve an entry, denoted \(\text{DB}[i]\), without the server knowing which specific entry is being requested. The concept of PIR was first introduced by Chor, Goldreich, and Kushilevitz~\cite{chor1995}, and since then, it has found numerous applications,  including private contact discovery \cite{contactdiscovery} and private DNS queries \cite{DNS}.

Beimel et al.~\cite{beimel} were the first to demonstrate that if neither the client nor the server store extra bits of information, then the number of entries R read by the server has the following lower bound: $R = \Omega(n)$ where $n$ is the size of the database. To overcome this limitation, Corrigan-Gibbs and Kogan~\cite{Corrigan} proposed to introduce client-side preprocessing. In their approach, the client performs an expensive, offline, query-independent preprocessing phase to collect \textit{hints} that can later be used for efficient private queries that can be answered in sublinear server time. 

In this practical work, we focused on schemes that utilize \textit{preprocessing PIR} and rely solely on the \textit{One-Way Function} (OWF) primitive. The strength of the studied schemes lies mainly in their simplicity of assumptions and, consequently, their ease of implementation. Notably, they do not require \textit{Public-Key Cryptography}, unlike many other solutions \cite{Chang}\cite{Boneh}.

Specifically, we focused on three key papers: 

\begin{itemize}
    \item \textbf{TreePIR}~\cite{TreePIR}, which have $O(\log{n})$ communication in upload but requires the assumption of two non-colluding servers.
    \item \textbf{PIANO}~\cite{Piano}, which uses a single server but has a bandwidth complexity of \( O(\sqrt{n}) \).
    \item \textbf{PPPS}~\cite{Efficient}, which combines ideas from PIANO and TreePIR to design a single-server scheme with a bandwidth complexity of \( O(n^{1/4}) \).
\end{itemize}

After establishing the mechanisms that allowed PIANO and PPPS to propose efficient 1-server schemes, we attempted to adapt them to TreePIR, which had the best upload bandwidth but was originally a 2-server scheme. By introducing a hint structure with two tables (primary table, backup table), inspired by have been done in PIANO, and by managing the refreshing of these hints using a technique called resampling, introduced in \cite{Efficient}, we succeeded in effectively transforming it into a single-server scheme with logarithmic bandwidth for upload and $O(\sqrt{n}\log{n})$ for download. The scheme still needs the client to support an additional storage of $O(\sqrt{n}\log{n})$. In terms of upload bandwidth, we achieve better performance than Shi et al.~\cite{Efficient}, whose bandwidth is in \( O(n^{1/4}) \), and PIANO, which is in \( O(\sqrt{n}) \). 
\section{Related Work}

\subsection{Performance Metrics and Concepts}
\paragraph{Relevant Metrics}
For PIR pre-processing, there are three different types of metrics relevant for performance comparison: 
\begin{itemize}
    \item \textbf{Storage:} Client or Server storage represents the additional storage (compared to a non-private data retrieval), required on the client/server side to execute the protocol.
    \item \textbf{Bandwidth:} The amount of data that needs to be transferred between the server and the client.
    \item \textbf{Computation:} Time needed for the client or server to generate the query, respectively the answer.
\end{itemize}

\paragraph{Global Concepts} 
The three schemes share similarities in their construction and flow; therefore, we will present them using the same framework:

The database is first split into \(\sqrt{n}\) chunks, each of size \(\sqrt{n}\). 
Then, in the preprocessing phase, the client generates hints whose main component consists of two parts: the first part is a set of \(\sqrt{n}\) indices, where each index comes from a different chunk of the database (hence, we have \(\sqrt{n}\) indices coming from \(\sqrt{n}\) different chunks). The second part is the parity of the values associated with this set of indices. \\
Then, in the query phase, when the client wants to retrieve the value associated with an index \( x \), it first select a hint that contains \( x \).  \\
This hint, which serves as the basis for the query sent to the server, is modified to reveal no information about \( x \): either \( x \) is simply removed from the query, or it is replaced with another value. The client then sends this modified set of indices.\\
Upon receiving this query, the server performs a computation and returns one or more additional parity values associated with the query. \\
With this response and the parity value associated with the hint, the client is then able to determine the value corresponding to index \( x \). \\
Finally, for security reasons, the client must replenish the hint table to avoid skewing the hint distribution. In the case of a two-server scheme, this is done by requesting a new hint containing \( x \) from the second server. In a one-server scheme, different techniques can be used, which we will describe later.

To explain the specificities of each scheme, we will follow a structure derived from the previously mentioned framework.  
First, we will detail the hint structure, followed by how the client obfuscates the target index before sending a query.  
Next, we will describe the computations performed by the server to generate a response, then the process the client follows to reconstruct the retrieved value.  
Finally, we will explain how the client replenishes its hints to maintain privacy.

\subsection{PIANO}
\textbf{PIANO} is the simplest of the three presented schemes. The authors' main contribution was to introduce a 3-table hint structure which allows them to only communicate with one server and to have a fast hint refreshing phase. 
\paragraph{Hint Structure}  
Hints are stored in three distinct tables. 

\begin{itemize}
    \item \textbf{Primary Table:} Contains entries of the form $(S, \bigoplus_{x \in S}{DB[x]})$, where S is a set of $\sqrt{n}$ indices where every index comes from a different chunk.

    \item \textbf{Replacement Table:} Contains entries of the form $(r, \text{DB}[r])$, where $r$ is a random index in the database and $DB[r]$ its associated value in the database.

    \item \textbf{Backup Table:} Each chunk of size $\sqrt{n}$ has its own backup entries. For each chunk $i$, it contains entries of the form $\{S, \text{DB}[S], y, \text{DB}[y]\}$, where $S$ is the same as in the primary table and $y$ is the value in $S$ corresponding to chunk $i$.  
\end{itemize}  
Using only pseudo-random functions (PRF), the authors managed to efficiently represent the hints. The set $S$ is actually stored using a PRF key $sk$, and whenever it need to be expanded, they use the following formula : $ S = \{i \cdot \sqrt{n} + \text{PRF}(\text{sk},i) | i \in [\sqrt{n}]\}$.

Therefore the hints only occupy $O(\sqrt{n}\log{\kappa} \alpha(\kappa))$ space and allows the client to answer $\sqrt{n} \log{\kappa} \alpha(\kappa)$ queries.  

\paragraph{Obfuscation}  
To retrieve an entry $x$, the client first searches for a primary entry $(S,\bigoplus_{y \in S}{DB[y]} )$ such that $x \in S$. Then, it retrieves an entry $(r, \text{DB}[r])$ from the replacement table, where $r$ belongs to the same chunk as $x$. Instead of directly querying $x$, the client constructs the query $Q = S \setminus \{x\} \cup \{r\}$,  With this modification, the server is unable to tell which index the client wants to retrieve. The client sends $\sqrt{n}$ values.

\paragraph{Server Computation}  
Upon receiving the query $Q$, the server computes the XOR of all elements indexed in $Q$: 
\[
R = \bigoplus_{y \in Q} {DB[y]}
\]
The response $R$ is sent back to the client. 
Since the server must compute $R$ from $\sqrt{n}$ values, the server computation requires about $O(\sqrt{n})$ operations. 

\paragraph{Client Reconstruction}  
Once the response $R$ is received, the client reconstructs the value at index $x$ by computing:  
\[
DB[x] =
\left(\bigoplus_{z \in S}{DB[z]} \right) \oplus \text{DB}[r] \oplus R
\]
This operation effectively cancels out the obfuscation introduced in the query, allowing the client to recover $\text{DB}[x]$ without leaking information about its true request to the server.  

\paragraph{Hint Refreshing}  
To maintain security, the primary hints must be continuously refreshed to prevent statistical biases in the queries. Since a hint containing $x$ was used, it must be replaced by another hint that also contains $x$. The client will take one of the entry for $x$'s chunk in the backup table, change the concerned index with $x$ and modify the parity value accordingly. 

\paragraph{Strengths:}
The primary advantage of this scheme lies in its simplicity. Indeed, the only use of one-way functions is for compacting the primary entries and enabling fast membership testing. 

\paragraph{Weaknesses:}
The main downside of this scheme is its bandwidth, both in upload and download, which is in $O(\sqrt{n})$, higher compared to the other schemes that will be detailed later.

\subsection{TreePIR}
TreePIR introduces a new way of storing and representing hints, which induces a gain in upload communication: the client only needs to transfer $O(\log{n})$ values.
\paragraph{Hint Structure}  

TreePIR leverages a double-lengthening PRF $G$, where:  
\[
k \rightarrow G(k) = G_0(k) \| G_1(k)
\]
with $|G_0(k)| = |G_1(k)| = |k|$. This structure allows for the efficient representation of a set of indices using a binary tree, where leaves correspond to hint values. A given index can be encoded as its position in the tree concatenated with its value, reducing the storage required for hints. Below is an illustration of a toy example with a database with 8 chunks. For example, the index for the $3^{rd}$ chunk would be : $010||G_{010}(k)$ 
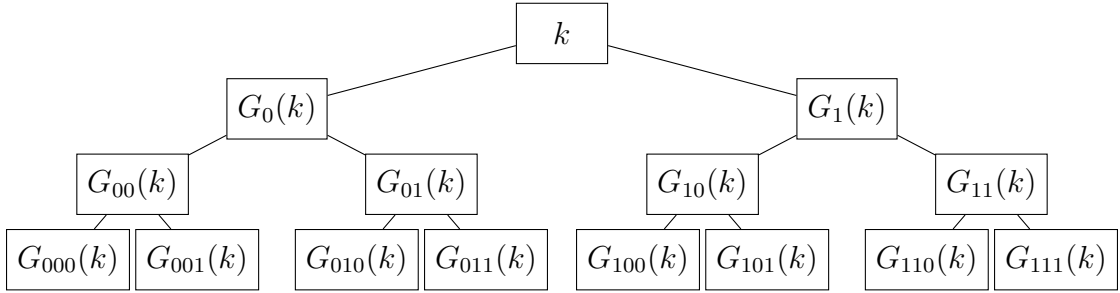
\begin{center}
\begin{tikzpicture}[
  level distance=1cm,
  level 1/.style={sibling distance=7.5cm},
  level 2/.style={sibling distance=3.8cm},
  level 3/.style={sibling distance=1.7cm},
  every node/.style={draw, rectangle, minimum width=1.2cm, minimum height=0.8cm, align=center}
]
  \node {$k$}
    child { node {$G_0(k)$}
    child { node {$G_{00}(k)$}
        child { node {$G_{000}(k)$} }
        child { node {$G_{001}(k)$} }
      }
      child { node {$G_{01}(k)$}
        child { node {$G_{010}(k)$} }
        child { node {$G_{011}(k)$} }
      }
    }
    child { node {$G_1(k)$}
      child { node {$G_{10}(k)$}
        child { node {$G_{100}(k)$} }
        child { node {$G_{101}(k)$} }
      }
      child { node {$G_{11}(k)$}
        child { node {$G_{110}(k)$} }
        child { node {$G_{111}(k)$} }
      }
    };
\end{tikzpicture}
\end{center}

\paragraph{Obfuscation}  
When the client wants to retrieve the value corresponding to an index $x$, it first searches for a hint containing $x$ and "punctures" the key accordingly. For example, if $x$ is located in chunk $001$, the client sends the keys $G_1(k)$, $G_{01}(k)$ and $G_{000}(k)$ and to the server in that order. The act of puncturing ensures that, with the given keys, the server cannot determine which chunk is missing or what the masked value is. However, it will still be able to compute all the other values. This guarantees privacy while minimizing communication overhead.

\paragraph{Server Computation}  
When the server receives the keys, it only knows that there is one key per tree level but it cannot establish their position in the three. This leads to $\sqrt{n}$ possible partial trees, each missing exactly one value due to the puncturing process. The server then computes the associated parities, for each tree. This takes $O(\sqrt{n}(\log{n})^2)$ time, because you have to build the trees and compute the parities.  The original TreePIR paper proposes two methods to transmit those parities. 
 
\begin{itemize}
    \item The problem effectively reduces from a PIR on a database of size $n$ to a PIR on an array of size $\sqrt{n}$. Since TreePIR cannot be reused dynamically, other PIR schemes, such as those based on Learning With Errors (LWE) \cite{LWE}, can be used to retrieve the data without preprocessing.
    \item Alternatively, the client can increase download bandwidth to $\sqrt{n}$ by requesting the entire array of parities.
\end{itemize}  

\paragraph{Client Reconstruction}  
From the array of parities, the client can find which value it needs to compute $DB[x]$, then this XOR value will be the same xor that the client had in its hint, except that the $DB[x]$ value is missing. Then if the parity value of the server is R and the parity value of the hint is P we have: 
\[ DB[x] = R \oplus P\]

\paragraph{Hint Refreshing}  
Since TreePIR is a two-server scheme, the client sends a query to the second server with a fresh key containing $x$ and puncturing it in the same way. With the parity value received from the second server and $DB[x]$ recovered from the first server, the client can construct a new fresh hint. 

\paragraph{Strengths}
TreePIR achieves sublinear upload bandwidth while maintaining
a client storage of $O(\sqrt{n})$ comparable to the other schemes.
\paragraph{Weaknesses}
TreePIR introduces additional computational costs. Specifically, client-side computation increases to $O(\sqrt{n} \log{n})$, while the server must perform $O(\sqrt{n} (\log{n})^2)$ operations. Also, without allowing other cryptographic primitives, the download bandwidth remains in $O(\sqrt{n})$.
\subsection{Efficient Pre-processing PIR Without Public-Key Cryptography}
In this paper, Shi et al. \cite{Efficient} introduce an important concept: Privately Programmable Pseudorandom Set with List Decoding (PPPS) which allows them to decrease the communication to $O(n^{1/4})$.
\paragraph{Hint Structure}  
The hints are composed of two tables: 
\begin{itemize}
    \item A hint table storing sets of values, with one value per chunk, and the associated parity value.  
    \item A replacement table, as in PIANO, filled with tuples: $(r,DB[r])$,  where $r$ is an index and $DB[r]$ its associated value. 
\end{itemize}  
Although the sets in the hint table always contain one value per chunk, PPPS introduces a new structure: values are grouped into superblocks of size \( n^{1/4} \). A master key derives \( n^{1/4} \) subkeys, each responsible for managing the $n^{1/4}$ values within its corresponding superblock.

\paragraph{Obfuscation}  
To retrieve an index $x$, the client first selects a hint containing $x$ and a replacement entry $(r, \text{DB}[r])$ where $r$ and $x$ come from the same chunk. Then it expands the master key into the \( n^{1/4} \) subkeys : $k_1,\dots,k_{n^{1/4}}$. Then it also expands the key $k_i$ in charge of $x$'s superblock into $n^{1/4}$ offsets (the actual indices) : $\delta_1,\dots,\delta_{n^{1/4}}$  
Finally it will send to the server the following : 
\[
Q = \{k_1,\dots,\tilde{k},\dots,k_{n^{1/4}},\delta_1,\dots,\tilde{\delta},\dots,\delta_{n^{1/4}}\} 
\]
where $\tilde{k}$ is a fresh key that replaces the key $k_i$, and $\tilde{\delta}$ is actually $r \mod \sqrt{n}$, the offset for the index $r$.  
By sending this, there is no trace of $x$ in the query, and the server cannot determine which key has been changed or which offset has been modified.

\paragraph{Server Computation}  
After receiving the obfuscated hint, the server computes $n^{1/4}$ parity values as follows: it constructs the set of values by expanding all the keys except one, which is directly replaced by the transmitted offsets.  
Moreover, since the parity values are almost entirely composed of the same values (differing only by one superblock), they can be efficiently computed in $O(\sqrt{n})$ operations.
 
\paragraph{Client Reconstruction}
With these $n^{1/4}$ parity values, the client can retrieve the one where the offsets are placed correctly (where the key $\tilde{k}$ is not used). With this parity value, the value is recovered as it is done in PIANO. 
\paragraph{Hint Refreshing}  
PPPS supports both one-server and two-server configurations. In a two-server model, one server processes queries while the other refreshes hints, assuming no collusion. In a one-server setting, one of the two mechanisms can be employed to prevent hint distribution skew:  
\begin{itemize}
    \item \textbf{Broken Hints :} The client ensures $x$ remains in the hint set by sampling a new "broken" superkey where the parity value of the set is marked unknown. However, for every $x$ requested, several queries must be sent : a "correct hint" must be sent for correctness, but many "broken" ones also has to be sent to preserve privacy. 
    \item \textbf{Resampling :} This technique introduces a backup table as it is done in PIANO. However the refreshing is a bit more complex : when a backup entry is promoted to the primary table, it is marked with a constraint $+x$ and when this key is used for another query, then the $x$'s superblock key is resampled to ensure that it contains $x$. 
\end{itemize}  

\paragraph{Strengths}  
 With PPPS, the authors achieve sublinear bandwidth ($O(n^{1/4})$) in a one-server setting.
\paragraph{Weaknesses}  
Even though the authors achieve better overall performance than in PIANO, their upload bandwidth is still greater than the one achieved by TreePIR.

\subsection{Performance comparison}
The tables \ref{tab:online} and \ref{tab:offline} sum up the online and offline performance of the discussed protocols. We named \textbf{PPPS1} the 1-server scheme from \cite{Efficient} where broken hints are used and \textbf{PPPS2} the one where the resampling technique is used. 
In the following, $\lambda$ is a computational security parameter and $\kappa$ is a statistical security parameter.
\begin{table}[h!]
    \centering
    \caption{Online Performance Comparison}
    \label{tab:online}
    \renewcommand{\arraystretch}{1.1} 
    \setlength{\tabcolsep}{6pt} 
    \begin{tabular}{@{} l p{2.8cm} p{3.5cm} p{2.8cm} p{3.0cm} c @{}}
        \toprule
        \textbf{Scheme} & \textbf{Storage} & \textbf{Bandwidth (Up/Down)} & \textbf{Client Time} & \textbf{Server Time} & \textbf{\# Servers} \\
        \midrule
        \textbf{TreePIR}  & $O_{\lambda}(\sqrt{n})$ & $O_{\lambda}(\log{n})/O(\sqrt{n})$ & $O(\sqrt{n}\log{n})$ & $O(\sqrt{n}\log^2(n))$ & 2 \\
        \textbf{PIANO}    & $O_{\lambda}(\sqrt{n}\log{\kappa}\alpha(\kappa))$ & $O(\sqrt{n})$ & $O_{\lambda}(\sqrt{n})$ & $O(\sqrt{n})$ & 1 \\
        \textbf{PPPS 1}   & $O_{\lambda}(\sqrt{n}\log{\kappa}\alpha(\kappa))$ & $O_{\lambda}(n^{1/4}\log{\kappa}\alpha(\kappa))$ & $O_{\lambda}(\sqrt{n}\log{\kappa}\alpha(\kappa))$ & $O_{\lambda}(\sqrt{n}\log{\kappa}\alpha(\kappa))$ & 1 \\
        \textbf{PPPS 2}   & $O_{\lambda}(\sqrt{n}\log{\kappa}\alpha(\kappa))$ & $O(n^{1/4})$ & $O_{\lambda}(\sqrt{n})$ & $O_{\lambda}(\sqrt{n})$ & 1 \\
        \bottomrule
    \end{tabular}
\end{table}

\begin{table}[h!]
    \centering
    \caption{Offline Preprocessing Costs}
    \label{tab:offline}
    \begin{tabular}{@{}lccc@{}}
        \toprule
        \textbf{Scheme} & \textbf{Client Time} & \textbf{Server Time} & \textbf{Communication} \\
        \midrule
        \textbf{TreePIR} & $O_{\lambda}(\sqrt{n})$ & $O_{\lambda}(n\log{n})$ & $O_{\lambda}(\sqrt{n})$ \\
        \textbf{PIANO} & $O_{\lambda}(n\log{\kappa}\alpha(\kappa))$ & $O(n)$ & $O(n)$ \\
        \textbf{PPPS 1} & $O_{\lambda}(n\log{\kappa}\alpha(\kappa))$ & $O(n)$ & $O(n)$ \\
        \textbf{PPPS 2} & $O_{\lambda}(\sqrt{n}\log{\kappa}\alpha(\kappa))$ & $O_{\lambda}(n\log{\kappa}\alpha(\kappa))$ & $O_{\lambda}(\sqrt{n}\log{\kappa}\alpha(\kappa))$ \\
        \bottomrule
    \end{tabular}
\end{table}
\newpage
\section{Preliminaries}
\subsection{Definitions}

We begin by formally defining a single-server PIR scheme with preprocessing. We adopt the notions introduced in \cite{Efficient} and \cite{Piano}.
A single-server PIR scheme with preprocessing consists of two phases:

\begin{itemize}
    \item \textbf{Offline Setup Phase:} The client starts with no prior information, while the server holds a database $\text{DB}$ consisting of $n$ entries. For simplicity, we assume that each entry is a single bit. During this phase, the client interacts with the server to generate hints, which will be used in the next phase.
    
    \item \textbf{Online Phase:} This phase can be executed as many times as needed. When the client receives an index $x$, it sends a query to the server. The server then responds with an answer that enables the client to recover the value corresponding to index $x$.
\end{itemize}

A PIR scheme has to respect the following properties: 

\paragraph{Correctness}
Given a database $\text{DB}$ with entries indexed by $0, 1, \dots, n - 1$, a PIR scheme satisfies correctness if, for any queried index $x \in \{0, 1, \dots, n - 1\}$, the client successfully retrieves $\text{DB}[x]$, the $x$-th bit of the database.

Formally, correctness requires that for any security parameter $\lambda \in \mathbb{N}$, and for any $n$, and for $q$ polynomially bounded in $\lambda$, there exists a negligible function $\text{negl}$ such that for any database $\text{DB} \in \{0,1\}^n$, and any sequence of queries $x_1, x_2, \dots, x_q \in \{0, 1, \dots, n - 1\}$, an honest execution of the PIR scheme with $\text{DB}$ and queries $x_1, x_2, \dots, x_q$ returns the correct answers with probability at least $1 - \text{negl}(\lambda)$.

\paragraph{Privacy}
A single-server PIR scheme satisfies privacy if and only if there exists a probabilistic polynomial-time simulator $\text{Sim}(1^\lambda, n)$ such that for any probabilistic polynomial-time adversary $\mathcal{A}$ acting as the server, for any $n$ and $q$ polynomially bounded by $\lambda$, and for any database $\text{DB} \in \{0,1\}^n$, the adversary's views in the following two experiments are computationally indistinguishable:

\begin{itemize}
    \item \textbf{Real:} An honest client interacts with $\mathcal{A}(1^\lambda, n, \text{DB})$, where $\mathcal{A}$ acts as the server and may deviate arbitrarily from the prescribed protocol. At each query step $t \in [q]$, $\mathcal{A}$ adaptively selects the next query $x_t \in \{0, 1, \dots, n - 1\}$ for the client, and the client is invoked with $x_t$ as input.

    \item \textbf{Ideal:} A simulated client $\text{Sim}(1^\lambda, n)$ interacts with $\mathcal{A}(1^\lambda, n, \text{DB})$, which again acts as the server and may arbitrarily deviate from the prescribed protocol. At each query step $t \in [q]$, $\mathcal{A}$ adaptively selects the next query $x_t \in \{0, 1, \dots, n - 1\}$ for the client, but this time the client is invoked without receiving $x_t$ as input, the client's answer is completely independent from the query. 
\end{itemize}
\subsection{Weak Privately Puncurable PRF}
For our work, we have reused the structure from \cite{TreePIR}. All security, privacy, and correctness proofs have been established in this paper. Here, we will only provide a definition of this structure and a practical implementation. In this subsection and the next, \( n \) refers to the entry parameter of the wpPRF and is unrelated to any database.  

\subsubsection{wpPRF formal definition}
\paragraph{Definition}
A \textbf{Weak Privately Puncturable Pseudorandom Function} (wpPRF) $\mathcal{F}$ consists of a tuple of four algorithms:

\begin{itemize}
    \item \textbf{Gen}$(1^\lambda) \rightarrow k$:  
    Takes a security parameter $\lambda$ and returns a wpPRF key $k \in \{0,1\}^\lambda$.

    \item \textbf{Eval}$(k, x) \rightarrow y$:  
    Given $x \in \{0,1\}^n$, outputs an evaluation on key $k$ at $x$, resulting in $y \in \{0,1\}^m$.

    \item \textbf{Puncture}$(k, i) \rightarrow k_i$:  
    Given a wpPRF key $k$ and an input $i$ from the domain, outputs a punctured key $k_i$ at point $i$.

    \item \textbf{PEval}$(k_i, j, x) \rightarrow y$:  
    Given a punctured key $k_i$, a guessed punctured index $j$, and an evaluation point $x$, outputs the evaluation of $x$ under $k_i$ assuming the punctured index is $j$.
\end{itemize}

\paragraph{Security Properties}
We reproduce here the properties stated in treePIR\cite{TreePIR} that are expected from a wpPRF. \\
The $P.Eval$ function has to be a PRF: 
\paragraph{Pseudorandom Function (PRF)}
A PRF $F : \{0, 1\}^\lambda \times \{0, 1\}^n \to \{0, 1\}^m$ satisfies security if, for any $k \in \{0, 1\}^\lambda$ sampled uniformly at random, for any function $H$ sampled uniformly at random from the set of functions mapping $\{0, 1\}^n \to \{0, 1\}^m$, for any PPT adversary $\mathcal{A}$, there exists a negligible function $\nu(\lambda)$ such that:

\[
\left| \Pr\left[\mathcal{A}^{O_H(\cdot)} \to 1\right] - \Pr\left[\mathcal{A}^{O_{F}(k,\cdot)} \to 1\right] \right| \leq \nu(\lambda).
\]

\paragraph{Security in Puncturing}
A wpPRF $(\text{Gen}, \text{Eval}, \text{Puncture}, \text{PEval})$ satisfies security in puncturing if for $r \in \{0, 1\}^m$ sampled uniformly, $k \leftarrow \text{Gen}(1^\lambda)$, there exists a negligible function $\nu(\lambda)$ such that for any PPT adversary $\mathcal{A}$, $\mathcal{A}$ cannot distinguish between the following experiments with probability greater than $\nu(\lambda)$:

- \textbf{Expt$_0$}:  
  $x \leftarrow \mathcal{A}(1^\lambda)$, $\text{Puncture}(k, x) \to k_x$, $b' \leftarrow \mathcal{A}(k_x, \text{Eval}(k,x))$.
  
- \textbf{Expt$_1$}:  
  $x \leftarrow \mathcal{A}(1^\lambda)$, $\text{Puncture}(k, x) \to k_x$, $b' \leftarrow \mathcal{A}(k_x, r)$.

The security in puncturing guarantees that the puncturing reveals nothing about the value punctured.
\paragraph{Privacy in Puncturing}
A Weak Privately Puncturable PRF $(\text{Gen}, \text{Eval}, \text{Puncture}, \text{PEval})$ satisfies privacy in puncturing if, given a uniformly random $b \in \{0, 1\}$ and $k \in \{0, 1\}^\lambda$, there exists a negligible function $\nu(\lambda)$ such that for any probabilistic polynomial-time adversary $\mathcal{A}$, $\mathcal{A}$ cannot correctly guess $b$ with probability greater than $\frac{1}{2} + \nu(\lambda)$ in the following experiment:

- $k \leftarrow \text{Gen}(1^\lambda)$.

- $(x_0, x_1) \leftarrow \mathcal{A}(1^\lambda)$.

- $k_{x_b} \leftarrow \text{Puncture}(k, x_b)$.

- $b' \leftarrow \mathcal{A}(k_{x_b})$.

The privacy in puncturing guarantees that the puncturing reveals nothing about the point that was punctured.

\paragraph{Weak Correctness}
A Weak Privately Puncturable PRF $(\text{Gen}, \text{Eval}, \text{Puncture}, \text{PEval})$ satisfies weak correctness in private puncturing if, given $k \leftarrow \text{Gen}(1^\lambda)$, for any point $x \in \{0, 1\}^n$, $k_x \leftarrow \text{Puncture}(k, x)$, the following holds:

\[
\forall x' \in \{0, 1\}^n, x' \neq x, \quad \text{Eval}(k, x') = \text{PEval}(k_x, x, x').
\]
\subsubsection{Practical Implementation}

We now describe a practical implementation of \textit{wpPRFs}, which has also been explained in TreePIR\cite{TreePIR}. It relies on a double-lengthening PRF $G$:

\[
k \rightarrow G(k) = G_0(k) \| G_1(k)
\]
where $|G_0(k)| = |G_1(k)| = |k|$, which allows the function $G$ to be reused on its outputs. 
This function $G$ is used to define a binary tree structure. We define \( G_x(k) \) as the consecutive application of the function \( G \) according to the digits of \( x \). For example, if \( x = 110 \), then:

\[
G_x(k) = G_0(G_1(G_1(k)))
\]
For $n$ in $[0, 2^3]$, the tree is structured as follows:

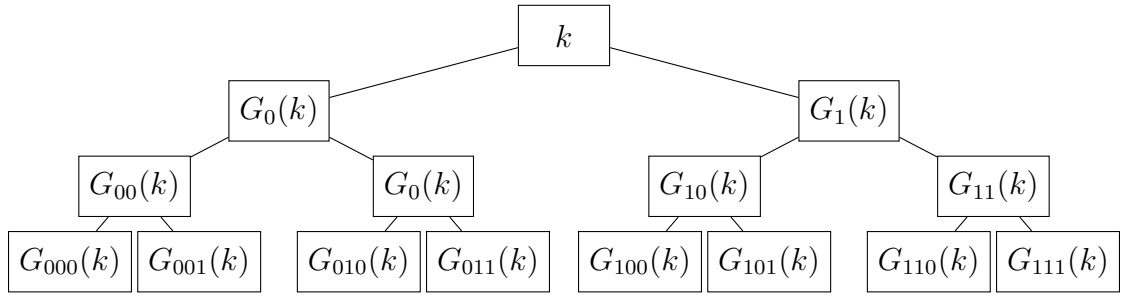
\begin{center}
\begin{tikzpicture}[
  level distance=1cm,
  level 1/.style={sibling distance=7.5cm},
  level 2/.style={sibling distance=3.8cm},
  level 3/.style={sibling distance=1.7cm},
  every node/.style={draw, rectangle, minimum width=1.2cm, minimum height=0.8cm, align=center},
  every label/.style={draw=none, rectangle, minimum width=0pt, minimum height=1pt, font=\small}
]
  \node {$k$}
    child { node {$G_0(k)$}
      child { node {$G_{00}(k)$}
        child { node {$G_{000}(k)$}}
        child { node {$G_{001}(k)$} }
      }
      child { node {$G_{0}(k)$}
        child { node {$G_{010}(k)$} }
        child { node {$G_{011}(k)$} }
      }
    }
    child { node {$G_1(k)$}
      child { node {$G_{10}(k)$}
        child { node {$G_{100}(k)$} }
        child { node {$G_{101}(k)$} }
      }
      child { node {$G_{11}(k)$}
        child { node {$G_{110}(k)$} }
        child { node {$G_{111}(k)$} }
      }
    };
\end{tikzpicture}
\end{center}

\paragraph{Puncture}
Suppose we want to puncture the index $100$.  
We highlight in red all subkeys that can be used to reconstruct the punctured value, therefore we transmit the remaining keys (in blue) to allow the reconstruction of the rest of the tree.

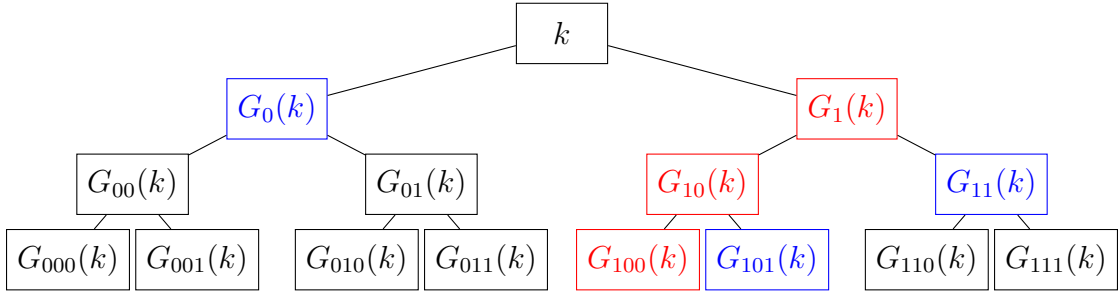
\begin{center}
\begin{tikzpicture}[
  level distance=1cm,
  level 1/.style={sibling distance=7.5cm},
  level 2/.style={sibling distance=3.8cm},
  level 3/.style={sibling distance=1.7cm},
  every node/.style={draw, rectangle, minimum width=1.2cm, minimum height=0.8cm, align=center}
]
  \node {$k$}
    child { node [blue]{$G_0(k)$}
      child { node {$G_{00}(k)$}
        child { node {$G_{000}(k)$} }
        child { node {$G_{001}(k)$} }
      }
      child { node {$G_{01}(k)$}
        child { node {$G_{010}(k)$} }
        child { node {$G_{011}(k)$} }
      }
    }
    child { node [red]{$G_1(k)$}
      child { node [red]{$G_{10}(k)$}
        child { node [red]{$G_{100}(k)$} }
        child { node [blue]{$G_{101}(k)$} }
      }
      child { node [blue]{$G_{11}(k)$}
        child { node {$G_{110}(k)$} }
        child { node {$G_{111}(k)$} }
      }
    };
\end{tikzpicture}
\end{center}

Thus, the punctured key consists of the concatenation of the three keys:  
\[
\left(G_{0}(k), G_{11}(k), G_{101}(k)\right)
\]
The order is always from the shallowest level to the deepest.

\paragraph{Punctured Evaluation (PEval)}
Given a punctured key and a point, we can reconstruct a tree with one missing value (the punctured one).

For instance, if we assume the puncturing happened at $000$ and we keep the punctured key from above, the reconstructed tree will appear as follows:

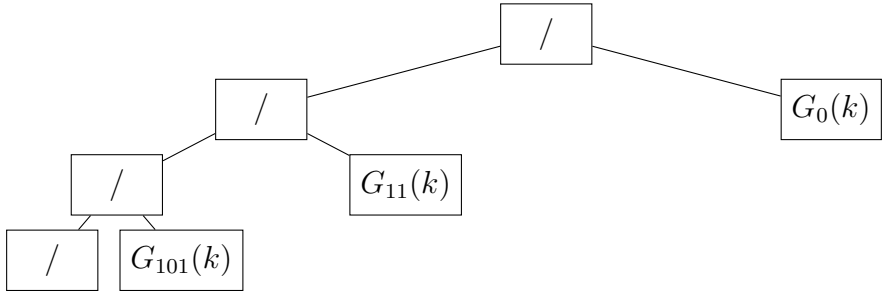
\begin{center}
\begin{tikzpicture}[
  level distance=1cm,
  level 1/.style={sibling distance=7.5cm},
  level 2/.style={sibling distance=3.8cm},
  level 3/.style={sibling distance=1.7cm},
  every node/.style={
    draw, rectangle, minimum width=1.2cm, minimum height=0.8cm, align=center
  }
]
  \node {$/$}
    child { node {$/$}
      child { node {$/$}
        child { node {$/$} }
        child { node {$G_{101}(k)$} }
      }
      child { node {$G_{11}(k)$} }
    }
    child { node {$G_0(k)$} };
\end{tikzpicture}
\end{center}

Since the value we used as a starting point to reconstruct the "tree $100$" was not the one punctured, the "tree $000$" is not correct: for example the value $G_{101}(k)$ should be $G_{001}(k)$. This does not challenge our definition of wpPRFs; we only require correctness when the starting point is the punctured point. \\
We reproduce in the following the formal implementation of a wpPRF as presented in TreePIR. In the following we use the notation $x_i$ to denote the first $i$ bits of $x$, and $|x_i|$ to denote the $i^{th}$ bit of $x$.
\begin{itemize}
        \item \textbf{Gen($1^\lambda$)}: 
            \begin{itemize}
                \item Outputs: A uniform string of length \(\lambda\).
            \end{itemize}
        
        \item \textbf{Eval(k, x)}: 
            \begin{itemize}
                \item Let \( y \leftarrow G_x(k) \). Output $y$
            \end{itemize}

        \item \textbf{Puncture(k, x)}: 
            \begin{itemize}
                \item Output list of seeds not in path to $x$, ordered by height. 
                \item Formally : output \( k_i = (sk_1, \dots, sk_{\log{n}} )\) , where \( sk_i = G_{p_i}(k) \) and \( p_i =x_{i-1} \parallel\overline{|x_{i}|} \).
            \end{itemize}

        \item \textbf{PEval($k_i, j, x$)}:
            \begin{itemize}
                \item Let \( y \leftarrow G_x(j, k_i) \), where \( G_x(j, k_i) \) denotes the leaf node at position \( x \) of the tree constructed from \( (j, k_i) \), where $j$ is a chunk and $k_i$ a punctured key.
            \end{itemize}
\end{itemize}
\subsection{Adding Resample to wpPRF}

This algorithm allows us to force the presence of a specific value in the tree from a punctured key. 
For instance, suppose we want to change the value at entry $111$ to a desired value $\beta$, in our previous toy example. 
From the $F.Puncture$ algorithm we had the key: $( G_{0}(k), G_{11}(k), G_{101}(k))$
Thus, in the reconstructed tree, the value at entry $111$ is handled by the key $G_{11}(k)$. Therefore, we will replace the key $G_{11}(k)$ with a different value: we will execute the $k' \leftarrow \{0,1\}^\lambda$ algorithm until we find a key $k'$ such that:
\[
G_1(k') = \beta.
\]
The new tree will now look like this: 
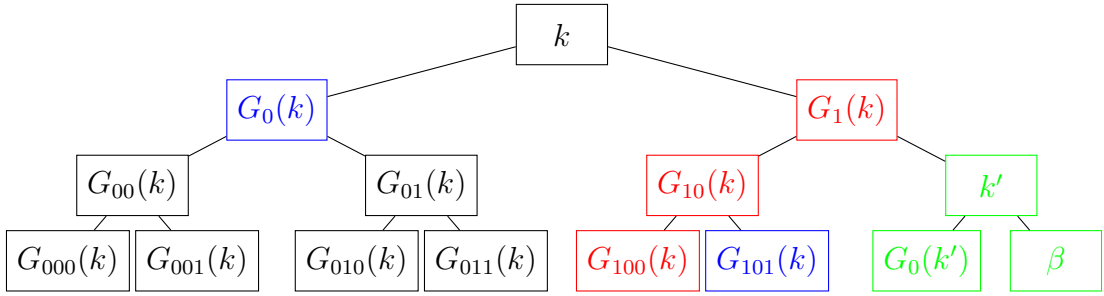
\begin{center}
\begin{tikzpicture}[
  level distance=1cm,
  level 1/.style={sibling distance=7.5cm},
  level 2/.style={sibling distance=3.8cm},
  level 3/.style={sibling distance=1.7cm},
  every node/.style={draw, rectangle, minimum width=1.2cm, minimum height=0.8cm, align=center}
]
  \node {$k$}
    child { node [blue]{$G_0(k)$}
      child { node {$G_{00}(k)$}
        child { node {$G_{000}(k)$} }
        child { node {$G_{001}(k)$} }
      }
      child { node {$G_{01}(k)$}
        child { node {$G_{010}(k)$} }
        child { node {$G_{011}(k)$} }
      }
    }
    child { node [red]{$G_1(k)$}
      child { node [red]{$G_{10}(k)$}
        child { node [red]{$G_{100}(k)$} }
        child { node [blue]{$G_{101}(k)$} }
      }
      child { node [green]{$k'$}
        child { node [green]{$G_{0}(k')$} }
        child { node [green]{$\beta$} }
      }
    };
\end{tikzpicture}
\end{center}
The new resampled key consists of:
\[
(G_0(k), k', G_{101}(k))
\]
Additionally, we can introduce negative constraints to exclude some specific values for any chunk affected by the resampling. For instance, if we want to exclude a value $z$ at entry $110$, the key $k'$ must satisfy:  
  \[
  G_1(k') = \beta \quad \text{and} \quad G_0(k') \neq z.
  \]

In the following, for an integer $x$, we define $x_l$ and $x_r$ as the left and right parts of its binary representation : if $|x| = 2t$ then $x_l$ are the leftmost $t$ bits of $x$'s binary representation, and $x_r$ the rightmost $t$ bits.  The resample algorithm takes as input a key, a punctured key, a positive constraint and possibly negative constraints, and can be formalized as follows: 
\begin{itemize}
        \item \textbf{Resample($k, k_{punc},+y, -c^1, \dots, -c^m$)}: 
            \begin{itemize}
                \item Parse $k_{punc}$ as $k_{punc} = (sk_1,\dots,sk_{\log{n}} ) $
                \item Find the subkey $sk_j$ such that : 
                \[
\exists p, p' \in \{1, \dots, n\} : G_p(k) = sk_j \quad\land \quad p \parallel p' = y_l
\]
                \item Sample a new key $sk_j' \leftarrow \{0,1\}^\lambda$ such that : \[
                G_{p'}(sk_j') = y_r
                \]
                \item Additionally, if there are some negative constraints $-c^1,\dots,-c^m$:\\ For any $c^i$ if there exists $p''$ such that $c^i_l$ can be written as $c^i_l = p || p'' $, then \( G_{p''}(sk'_j) \neq c^i_r \).
                    
                \item Outputs: \( [sk_1, \dots, sk'_j, \dots, sk_{\log{n}}] \).
            \end{itemize}
    \end{itemize}
\section{1 server - TreePIR}
In this part, $n$ is the size of the database, therefore the wpPRF structure will be used on a tree with $\sqrt{n}$ leaves : this is due to the fact that the database is split into $\sqrt{n}$ chunks of size $\sqrt{n}$.  
\subsection{Notations}
We make explicit some notations that will be needed for the protocol explanation: \\ 
\noindent\textit{Definition 1.} \quad For a key \( k  = F.\text{Gen}(1^\lambda) \), the set \( T(k) \) is defined as:
\[
T(k) = \{x_l ||  G_{x_l}(k) \mid x_l \in [1, \sqrt{n}] \}.
\]
This represents the leaves of the tree generated by the key \( k \). \\

\noindent\textit{Definition 2.} \quad For a (sub)key \( sk \) and a prefix $p \in \{0,1\}^{\leq\log{\sqrt{n}}}$, we define: \\
\[
\text{if } \text{len}(p) = {\log{\sqrt{n}}} :
\quad ST(sk, p) = \{ p \| sk \}
\]
\[
\text{else:} \quad
ST(sk, p) = 
\left\{ p \| x \| G_x(sk) \;\middle|\; x \in \left[ 0, \ldots, \frac{\sqrt{n}}{2^{\text{len}(p)}}-1 \right] \right\}.
\]
This represents the set of leaves handled by a subkey within a bigger tree.  \\ 

\noindent\textit{Definition 3.} \(\kappa\) denotes a statistical security parameter, \(\lambda\) denotes a computational security parameter. We use \(\alpha(\kappa)\) to denote an arbitrarily small super-constant function

\subsection{Protocol explanation}
\paragraph{Intuition}
Our proposed scheme is an adaptation of TreePIR to a single-server setting. It leverages a hint structure based on two tables, while incorporating the resampling technique from PPPS\cite{Efficient}. The key difference compared to PIANO is that when a hint from the backup table is marked to include a specific entry $y$, we cannot simply replace the value for $y$'s chunk with $y$ itself. This replacement was feasible in PIANO because the set is sent directly to the server. However, in our case, since we send a key that the server expands to retrieve the values for parity computation, our hints must include \textbf{constraints}. A \textbf{positive constraint} ensures that the set represented by the key contains a specific value, while \textbf{negative constraint} prevents a particular value from appearing in the set. Further details are provided in the following sections.
We will first give a full description of the protocol for $\sqrt{n}/2$ random, distinct queries and after this, we will discuss how the protocol can be extended.

\paragraph{Hint structure}
The client maintains two tables: the primary table and the backup table, both influenced by the resampling process.

The primary table has $M_1 = \sqrt n \log \kappa \cdot \alpha(\kappa)$ entries. Each entry is a triplet $(sk,p_{sk},(c^1,\dots,c^l))$:

\begin{itemize}
    \item $sk$  represents the secret key describing the tree structure.
    \item $p_{sk}$ represents the XOR-sum of every value in \( T(sk) \), formally defined as:
    \[
    p_{sk} = \bigoplus_{i \in T(sk)} DB[i]
    \]
    \item $(c^1, \dots, c^l)$ represents a tuple of constraints
\end{itemize}

For the backup table, for each chunk \( c \), we have \( M_2  = 3 \log \kappa \cdot \alpha(\kappa)\) entries. Each entry in the backup table is composed of the following three components:

\begin{itemize}
    \item $sk$, $ p_{sk}$ are defined as in the primary table
    \item A tuple containing the XOR-sum for every subtree that contains the chunk $c$, formally expressed as $(\gamma_1,\dots,\gamma_{\log{\sqrt{n}}})$ where : 
    \[
    \gamma_i =  \bigoplus_{l \in ST(G_{c_i}(sk), c_{i})} DB[l] \quad
    \]
    where $c_{i}$ represents the first $i$ bits of the binary representation of the chunk c. 
\end{itemize}
\paragraph{Hint Selection and Obfuscation}
When the client seeks the value corresponding to index $x$, it first locates a hint $(sk, p_{sk}, S)$ such that $P.\text{eval}(sk, x_l) = x_r$. Two scenarios arise: 
\begin{itemize}
    \item If there is no positive constraints on $S$, the client computes the punctured key and directly transmits it to the server.
    \item If there is a positive constraint and potentially negative constraints, the client first punctures the key and then executes the resampling algorithm to obtain a punctured key that satisfies all constraints and then sends it to the server. 
\end{itemize}
If the positive constraint is on $x$'s chunk, we ignore this hint. 
\paragraph{Server Computation}
With the received punctured key, the server proceeds as in TreePIR, computing the parity values for all trees it can reconstruct. The parity computation follows a left-to-right order based on the punctured chunk index, ensuring efficient processing by leveraging previously computed results.

For each potential punctured chunk, the server also returns the parity values for every repositioned key in the tree. Taking our toy example with 8 indices, suppose the punctured chunk is $100$. The server then returns the parities, $\beta_{100,1}$, $\beta_{100,2}$, and $\beta_{100,3}$, structured as follows:

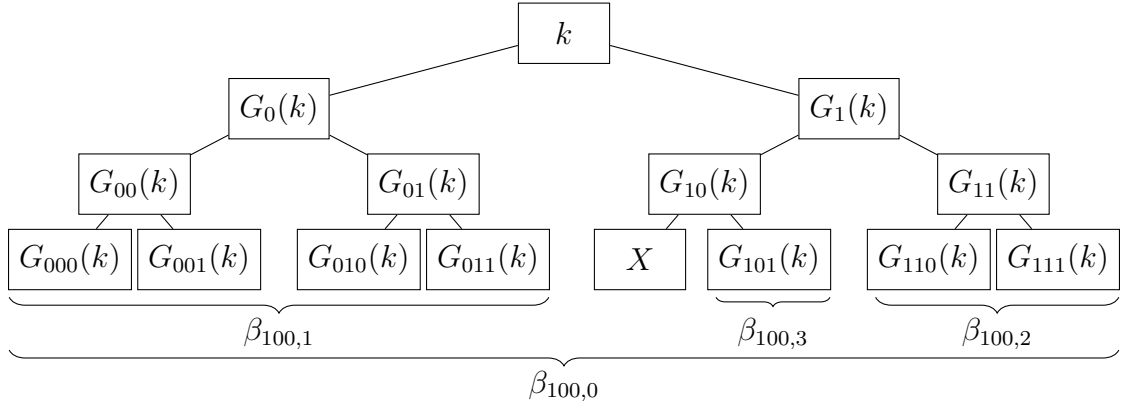
\begin{center}
\begin{tikzpicture}[
  level distance=1cm,
  level 1/.style={sibling distance=7.5cm},
  level 2/.style={sibling distance=3.8cm},
  level 3/.style={sibling distance=1.7cm},
  every node/.style={draw, rectangle, minimum width=1.2cm, minimum height=0.8cm, align=center}
]
  \node {$k$}
    child { node {$G_0(k)$}
      child { node {$G_{00}(k)$}
        child { node {$G_{000}(k)$} }
        child { node {$G_{001}(k)$} }
      }
      child { node {$G_{01}(k)$}
        child { node {$G_{010}(k)$} }
        child { node {$G_{011}(k)$} }
      }
    }
    child { node {$G_1(k)$}
      child { node {$G_{10}(k)$}
        child {node{$X$} }
        child { node {$G_{101}(k)$} }
      }
      child { node {$G_{11}(k)$}
        child { node {$G_{110}(k)$} }
        child { node {$G_{111}(k)$} }
      }
    };

  \draw [decorate,decoration={brace,amplitude=3.5pt,mirror}] (2,-3.5) -- (3.5,-3.5) node[midway,below=2pt,draw=none] {$\beta_{100,3}$};
  \draw [decorate,decoration={brace,amplitude=6pt,mirror}] (4.1,-3.5) -- (7.3,-3.5) node[midway,below=2pt,draw=none] {$\beta_{100,2}$};
  \draw [decorate,decoration={brace,amplitude=6pt,mirror}] (-7.3,-3.5) -- (-0.2,-3.5) node[midway,below=2pt,draw=none] {$\beta_{100,1}$};
  \draw [decorate,decoration={brace,amplitude=6pt,mirror}] (-7.3,-4.2) -- (7.3,-4.2) node[midway,below=2pt,draw=none] {$\beta_{100,0}$};

\end{tikzpicture}
\end{center}
$\beta_{100,0}$ is computed as the XOR of the other $\beta$ values. In principle, the server would not need to send it, as the client could reconstruct it using the other $\beta$ values. However, for the sake of clarity and notational simplicity, we choose to explicitly include it in our formulation.
\paragraph{Client computation}
Let $c$ be the chunk $x$ belongs to. Upon receiving the parity matrix, the client follows two possible paths:

If there were no initial constraints, the client simply computes:
  \[
  \text{DB}[x] = \beta_{c,0} \oplus p_{sk}
  \]
  using the notations defined in the hint structure.

If resampling is required, the computation becomes more complex. The client initially computed parities on the original tree, while the server computed them on the resampled tree. Nevertheless, the client can still recover $\text{DB}[x]$. Since entries from the backup table also store the parity for the resampled tree, and the server also returns parity values for this subtree, the final computation is:

  \[
  \text{DB}[x] = \beta_{c,0} \oplus p_{sk} \oplus \gamma_h \oplus \beta_{c,h}
  \]

  where the term $p_{sk} \oplus \gamma_h \oplus \beta_{c,h}$ ensures consistency between the precomputed hint and the response received from the server, and where $h$ is the distance to the root of the tree of the largest subtree that contains $y$ and not $x$.

\paragraph{Refreshing}
To refresh the hints, the client selects an entry from the backup table and adds the constraint $+y$ to its constraint tuple. Additionally, since the client checked for every preceding hints before the one chosen whether $x$ was present and found that it was not, it must mark the constraint $-x$ for those hints. This step is crucial: let's imagine that a query $z$ arises and triggers a resampling that affects $x$'s chunk then $x$ could appear, skewing its distribution and making it more likely to appear than other elements in its chunk. Such a bias could compromise the privacy of the server’s responses.

All phases described above are formally detailed in the following boxes: the first outlines the execution of a single query, while the second presents the full protocol.

\begin{mdframed}[linewidth=0.001\textwidth,frametitle = Detailed Protocol for One Query,frametitlealignment=\centering,]
\label{yoyo}
\small
\subsubsection*{Client's Input}
The client begins by providing a hint:
\begin{itemize}
    \item A master Tree key \( sk \), assuming \( F.Eval(k,x_{l}) =x_{r} \);
    \item The parity of the tree generated by \(sk\) assumed to be \( p_{sk} \);
    \item A set of constraints $c^1,\dots, c^m$
    \item if the hint is marked with a positive constraint \(+y\), it also provides the parity for every subtree containing $y$'s chunk : \[\forall i \in \{1,\dots,\log{\sqrt{n}}\}, \quad\gamma_i = \bigoplus_{l \in ST(k_i,y_i)}DB[l]\] where \[k_i = G_{y_i}(sk)\]
\end{itemize}

\subsubsection*{Step 1: (Client)}
The client performs the following operations:
\begin{enumerate}
    \item It executes : \(k_{0} = F.Puncture(sk,x)\) 
    \item \textbf{If} the hint comes from the backup table and marked with \(+y\), and possibly other negative constraints $-c^1, \dots, -c^m$ it executes : \\ \(k_{1} = F.Resample(sk,k_{0},+y,c^1,\dots,c^m)\). \\
    \textbf{else} \(k_1 = k_0\) 
    \item It sends \( k_1 \) to the server.
\end{enumerate}
\subsubsection*{Step 2: (Server)}
Upon receiving \( k \), the server performs the following actions:
\begin{enumerate}
    \item The server parses $k$ as $[k_1,\dots,k_{\log{\sqrt{n}}}]$
    \item For each possible chunk punctured $c$, the server will compute:\\ $\forall i \in [1,\log{\sqrt{n}}],$ \(\beta_{c,i} = \bigoplus_{z \in ST(k_i,p_{i})}\text{DB}[z] \text{ with } p_{i} = c_{i-1} + \overline{|c_{i}|}\)
    \item It also computes \(\beta_{c,0} = \bigoplus_{i=1}^{\log{\sqrt{n}}} \beta_{c,i}\), the XOR value of the whole tree.
    \item Finally, it returns the following matrix to the client: 
    \[
M =
\begin{pmatrix}
\beta_{1,0} & \beta_{2,0} & \ldots & \beta_{\sqrt{n},0} \\
\beta_{1,1} & \beta_{2,1} & \ldots & \beta_{\sqrt{n},1} \\
\vdots & \vdots & \vdots & \vdots\\
\beta_{1,\log{\sqrt{n}}} & \beta_{2,\log{\sqrt{n}}}& \ldots & \beta_{\sqrt{n},\log{\sqrt{n}}} \\
\end{pmatrix}
\]
\end{enumerate}
\small
\subsubsection*{Step 3: (Client)}
Upon receiving M, the client performs the following actions:
\begin{itemize}
    \item If the hint was coming directly from the primary table, saves $p_{sk}\oplus \beta_{i,0}$ as the answer where $i$ is $x$'s chunk.
    \item If the hint was a backup hint : Let $h$ be distance to the root of the largest subtree that contains $y$'s chunk and not $x$, then the client can compute : 
    \[
    DB[x] = \beta_{i,0} \oplus p_{sk} \oplus \gamma_h \oplus \beta_{i,h} 
    \]
\end{itemize}
\end{mdframed}
\begin{mdframed}[frametitle=Single Server Scheme for \(Q = \sqrt{n}/2\), frametitlealignment=\centering]
\small
    \subsubsection*{Preprocessing}
    The protocol is designed to operate in a streaming mode, where the database is processed in chunks, and the Primary Table and Backup Table are updated dynamically with each chunk received.

    \begin{enumerate}
            \item The client generates $M_1$ keys $sk_1,\dots,sk_{M_1}$ where $sk_i = F.Gen(1^\lambda)$ for the primary table, and $\sqrt{n} \cdot M_2$ keys for the backup table.
            \item As each chunk is processed, the client can compute the values needed in each hint : 
            When the client receives the $j-th$ chunk : 
            \begin{itemize}
                \item Primary table: update the $p_{sk}$ value: $p_{sk} \leftarrow p_{sk} \oplus DB[F.Eval(sk,j)]$
                \item Backup table: update the $p_{sk}$ and $\gamma$'s values.
            \end{itemize}
    \end{enumerate}
\subsubsection*{Online Query for index $x = x_l||x_r \in \{0,1,\dots,n-1\}$}
\begin{enumerate}
    \item Find the first entry in the primary table such that for the key $sk$ in this hint  $F.Eval(sk,x_{l}) = x_r$ 
    \item Execute the subroutine described before.
\end{enumerate}
\subsubsection*{Refreshing}
    \begin{enumerate}
        \item  Client replaces the matched hint with the first unconsumed hint from the  backup hint group for $x$'s chunk and mark it with $+x$.
        \item In the primary table, for every hint located before the used hint, mark the hint with $-x$.  
    \end{enumerate}
            
\end{mdframed}
\paragraph{Support unbounded, arbitrary queries}
For security, correctness and efficiency issues, we assumed that the queries were random and distinct and that we had only $\sqrt{n}/2$ queries.\\ 
As it is done in PIANO \cite{Piano}, we can get rid of the "distinct" assumption easily. We can require the client to store the answer for the most recent $\sqrt{n}/2$ queries . If a query is repeated, the client retrieves the answer locally and sends instead a random and independent query to the server to mask the duplication.

To eliminate the need for the random assumption, which is only required for load balancing across chunks, we assume that the server publishes a PRP key. This is used to shuffle the database indices. When the client wishes to retrieve an entry at a certain index, it queries the server for the corresponding permuted index, effectively randomizing the queries. As noted in PIANO, if the PRP key is not honestly generated, it does not impact privacy but may affect correctness, though correctness is inherently unverifiable if the server is malicious.

Finally, to support unbounded queries, we use a pipelining trick: we can require the server to send two chunks whenever it sends an answer to query. Since the download bandwidth is already $O(\sqrt{n}\log{n})$ sending two chunks along with the answer does not affect the asymptotic communication complexity. However it has an impact on the client computation, which is now: $O_\lambda(\sqrt{n}\log{n}\log(\kappa) \alpha(\kappa))$, the details for this calculation are provided in the section \ref{subsec:efficiency}. 

\subsection{Privacy Proof}
Let F be a $wpPRF$, augmented with a resampling algorithm. 
\subsubsection*{Ideal Game}
We define the following Ideal game:

\begin{itemize}
    \item \textbf{Offline}: The adversary $\mathcal{A}$ receives the streaming signal.
    \item \textbf{Online}: For any query, A chooses the query \( x^t \) to send to the client. The client ignores it, picks a random \( y \in [1,\dots,n] \) as its query index. The client samples a new key $k$ (constrained to $G_{y_l}(k) = y_r$) and executes the TreePIR protocol. The client then sends the punctured key.
\end{itemize}

This Ideal defines a PPT simulator unrelated to \( x^t \).

\subsubsection*{Hyb1}

\begin{itemize}
    \item \textbf{Offline}: A receives the streaming signal. The client samples $sk_1$, \dots, $sk_{M_1}$, and generates the corresponding hints: $(sk_i,0, ())$. 
    \item \textbf{Online}: For each round $t$ $A$ chooses a query $x^t$
    \begin{itemize}
    \item The client finds the first matched key $sk_i$ in the hint table, constrained to $G_{x^t_l}(sk_i) = x^t_r$
    \item The client computes $k_1 = F.Puncture(sk_i,x)$ and sends it to the server
    \end{itemize}
    \item \textbf{Refreshing}: the client replaces the entry $sk_i$ with a freshly-sampled key $sk'$ constrained to $G_{x^t_l}(sk') = x^t_r$

\end{itemize}

\paragraph{Proof.} The only difference between \textit{Ideal} and \textit{Hyb1} is inherent to the TreePIR protocol, so we did a proof similar to TreePIR\cite{TreePIR}

By the definition of the security of puncturing, a query to $x_l \| x_r$ is indistinguishable from a query to $x_l \| y_r$:
A punctured key reveals nothing about the evaluation at the punctured index. The keys $F.\text{Puncture}(k,x_r \| y_r)$ and $F.\text{Puncture}(k,x_l \| x_r)$ are computationally indistinguishable.
Moreover, by the definition of privacy in puncturing, the adversary cannot guess with probability greater than $\frac{1}{2} + \nu(\lambda)$, where $\nu(\lambda)$ is a negligible function of $\lambda$, which index was punctured. That is:
$
    F.\text{Puncture}(k,x_l \| y_r) \text{ and }F.\text{Puncture}(k,y_l \| y_r)
$ are computationally indistinguishable.

By the transitive property, the punctured keys for $x$ and $y$ are computationally indistinguishable. Since this is the only difference observed by the server, \textit{Ideal} and \textit{Hyb1} are indistinguishable for the server.

\subsubsection*{Hyb2}

\begin{itemize}
    \item \textbf{Offline}: $A$ receives the streaming signal. The client samples \( M_1 \) keys. It fills the primary table with hints of the form $(sk_i,0,())$
    $sk_i$ being a key, () being a placeholder for the future constraints.
    \item \textbf{Online}: For each round $t$, $A$ chooses a query $x^t$
    \begin{itemize}
    \item The clients finds the first matched key $sk_i$ in the hint table constrained to $G_{x^t_l}(sk_i) = x^t_r$
    \item If the hint is marked with the constraint $+y$, and possibly other negative constraints $-c^1,\dots,-c^p$, the client resamples a key $sk'$ subject to all constraints.
    \item The client computes $k_1 = F.Puncture(sk',x^t)$ and sends it to the server
    \end{itemize}
    \item\textbf{Refreshing}: 
    \begin{itemize}
        \item The client samples a new key $sk'$, and replaces the used hint with a new one : $(sk',{+x^t})$
        \item The client then marks the constraint \( -x \) for the previous \( i-1 \)-th hints
    \end{itemize}
\end{itemize}

\paragraph{Proof.} The main difference between the two hybrids is that in \emph{Hyb2}, we introduce constraints. We need to prove that the chosen hints and their subsequent sets follow the same distribution in both cases. Since the rest of the process is identical, the adversary's view will be the same, conditioned on the truth of the previous statement. 

We introduce a matched hint vector \( I = (i_1, \dots, i_t) \), where \( i_t \) represents the index of the chosen hint for each round. Note that since in \textit{Hyb2} the resampling step happens after the hint has been chosen in both hybrids, \( I \) follows the same distribution. We will prove that for every query $x^t$, and even if it receives the matched hint vector along with the query, it cannot distinguish between \textit{Hyb1} and \textit{Hyb2}. 

We will now introduce \textit{Expt1} and \textit{Expt2} to prove that in both experiments and conditioned on the same matched hint vector, the chosen hint for the \( t \)-th query follows the same distribution. \\

We define \textit{Expt1} as follows, for every query $x^t$:  
\begin{itemize}
    \item The client receives a vector of queries \( (x^1, \dots, x^t) \).
    \item The client samples the \( M_1 + t - 1 \) hints (conditioned on \( x^1, \dots, x^{t-1} \)) and generates the matched hint vector accordingly.
    \item The client then returns the matched hint vector and the punctured key.
\end{itemize}
The adversary cannot distinguish between \textit{Expt1} and \emph{Hyb1} queries at round $t$: in both cases, it just receives a punctured key, and the hints are distributed the same way in both experiments.

Let \( \alpha_I \) be the probability that the vector I is equal to some specific vector \( (i_1, \dots, i_{t-1}) \).  
We define \textit{Expt2} as follows:  
\begin{itemize}
    \item The client receives a vector of queries \( (x^1, \dots, x^t) \).
    \item The client samples a random vector \( I \), where the probability distribution is given by \( \alpha_I \) defined earlier.
    \item The client then samples the keys conditioned on this hint index vector.
    \item The client punctures the i-th query and sends it to the server.
\end{itemize}
By the definition of \( \alpha_I \), the expanded sets in the chosen hints for the \( t \) queries, in both \textit{Expt2} and \emph{Hyb2}, follow the same distribution (random sets conditioned on the same constraints).

A key observation in both cases (\textit{Expt1} and \textit{Expt2}) is that the chosen hints come from the posterior distribution (conditioned on \( \alpha_I \)). Therefore, from the adversary's perspective, the selected entries in both experiments share the same distribution.

\subsubsection*{Hyb3}

\begin{itemize}
    \item \textbf{Offline}: $A$ receives the streaming signal. The client samples \( M_1 \) keys. It fills the primary table with hints of the form $(sk_i,0,())$
    $sk_i$ being a key, () being a placeholder for the future constraints.
    \item \textbf{Online}: A chooses the query \( x^t \).
    \begin{itemize}
    \item The client finds the first key $sk_i$ such that $G_{x^t_l}(sk_i) = x^t_r$
    \item The client computes $k_1= F.Puncture(sk',x^t)$
        \item If the hint is marked with \( +y \) (and possibly other negative constraints $-c^1,\dots,-c^p$): 
        \begin{itemize}
            \item The client resamples a new key $sk'$ according to all constraints and also subject to $G_{x_l}(sk') = x_r$
            \item  The client computes
             $k_2 = F.Resample(sk',k_1,+y,-c^1,\dots,-c^p)$ and sends $k_2$ to the server
        \end{itemize}
        \item Else it sends $k_1$ to the server.
    \end{itemize}
    
    \item \textbf{Refreshing}: Same as in Hyb2 :  
    \begin{itemize}
    \item The client samples a new key $sk'$, and replace the used hint with a new one : $(sk',{+x^t})$
        \item The client then marks the constraint \( -x^t \) for the previous \( i-1 \)-th hints.
    \end{itemize}
\end{itemize}
\paragraph{Proof.} There is only one difference between \textit{Hyb2} and \textit{Hyb3} : the added resampling step. 
Since the rest of the protocol is the same, we only need to prove that the sets represented by the punctured keys in both hybrids come from the same distribution, and that the punctured keys appear identical in both hybrids. It establishes that \textit{Hyb2} and \textit{Hyb3} are indistinguishable.  \\
In \textit{Hyb2}, after the resampling step and due to the pseudorandomness of $G$,  we can say that the set of elements ("the leaves of the tree") comes from a random vector of independent variables \( (X_1, \dots, X_{\sqrt{n}} )\), where each random variable \( X_i \) is uniformly distributed over a set \( S_i \). This set \( S_i \) is almost identical to \( \{1, \dots, \sqrt{n}\} \), except that some values are excluded according to the following constraints: \\ \[
\forall c \in \{c^1, \dots, c^p\}, \, \text{if } c_l = i, \text{ then } c_r \notin S_i 
\]
\[
\text {if } x_l = i \text { then } S_i = \{x^t_r\} 
\]
\[
\text{if } y_l = i, \text{ then } S_i = \{y_r\}
\]

The algorithm \( F.\text{Resample} \) finds the subkey that represents the subtree containing the chunk of the positive constraint \( +y \), and resamples it accordingly to constraints on this subtree. This can be interpreted as selecting a subset of variables \( X_{j_1'}, \dots \), and resampling them according to the same constraints as the original variables \( X_{j_1},\dots \) that they replace. 

Ultimately, this results in a vector of pairwise independent random variables whose marginal distributions follow the same distribution as the original variables. Therefore, the set of elements represented by the punctured key is indistinguishable in both cases.

Furthermore, even if part of the punctured key is replaced by another element, the adversary cannot determine which part of the punctured key has been changed. This is because doing so would break the randomness of the key, which is not possible for a properly generated punctured key. Therefore the server cannot distinguish the punctured key obtained from \textit{Hyb3} and the one from \textit{Hyb4}.\\

We introduce \textit{Hyb4}. The difference with \textit{Hyb3} is that in \textit{Hyb4}, the first resampling step with all constraints does not take into account the positive constraint $+y$. This step is now performed only within the F.Resample algorithm.

\subsubsection*{Hyb4}
\begin{itemize}
    
\item \textbf{Offline :}  $A$ receives the streaming signal. The client samples \( M_1 \) keys. It fills the primary table with hints of the form $(sk_i,0,())$
    $sk_i$ being a key, () being a placeholder for the future constraints.
    \item \textbf{Online}: A chooses the query \( x_t \), The client finds the first key $sk_i$ such that $G_{x^t_l}(sk_i) = x^t_r$
    \begin{itemize} 
    \item The client computes $k_1= F.Puncture(sk',x^t)$ 
    \item If there is a positive constraint $+y$ :
    \begin{itemize}
        \item The client resamples the key according to all negative constraints $-c^1,\dots,-c^p$ and the constraint $+x^t$, but not the positive constraint $+y$.
        \item If there is a positive constraint $+y$, the client computes\\$k_2 = F.Resample(sk',k_1,+y,-c^1,\dots,-c^p)$ and sends $k_2$ to the server.
    \end{itemize}
    \item Else it sends $k_1$ to the server.
    \end{itemize}
     \item \textbf{Refreshing}: Same as in Hyb3 :  
     \begin{itemize}
        \item The client samples a new key $sk'$, and replaces the used hint with a new one : $(sk',{+x^t})$
        \item The client then marks the constraint \( -x^t \) for the \( i-1 \)-th hint.
    \end{itemize}
\end{itemize}

\paragraph{Proof} We maintain the formalism defined in the previous proof. 
The difference between the hybrids comes from the fact that in the original resampling step, the positive constraint $+y$ is no longer considered in \textit{Hyb4}. \\
Again, since the rest of the protocol is the same, we only need to show that the set of elements represented by the punctured keys in both hybrids comes from the same distribution. \\ 
In \textit{Hyb3}, all random variables $(X_1,\dots,X_{\sqrt n})$ are directly subject to the same constraints. Then, a subset of the variables is resampled, but with the exact same constraints as before. \\ 
In \textit{Hyb4}, the first resampling step only takes into account the following constraints:  
\[
\forall i \in \{1,\dots,\sqrt n\},\forall c \in \{c^1, \dots, c^p\}, \, \text c_l = i, \implies c_r \notin S_i 
\]
\[
\text {if } x^t_l = i \text { then } S_i = \{x^t_r\} 
\]

The second resampling step is performed in the \( F.\text{Resample} \) algorithm. 
The constraints can be represented as follows :
For every $X_i$ belonging to the subset of random variable that has to be resampled : 
\[
\forall c \in \{c^1, \dots, c^p\}, \, \text{if } c_l = i, \text{ then } c_r \notin S_i 
\] 
\[
\text{if } y_l = i, \text{ then } S_i = \{y_r\}
\]
In the end, the vector of random variables in both hybrids follows (pairwise) the same constraints in both hybrids, so the transmitted keys represent the same sets.

\subsubsection*{Real*}
Finally, we define Real*, where we remove the resampling of the key step: 
\begin{itemize}
\item \textbf{Offline :}  $A$ receives the streaming signal. The client samples \( M_1 \) keys. It fills the primary table with hints of the form $(sk_i,0,())$
    $sk_i$ being a key, $()$ being a placeholder for the future constraints.
    \item \textbf{Online}: A chooses the query \( x^t \), 
    \begin{itemize} 
    \item The client finds the first key $sk_i$ such that $G_{x^t_l}(sk_i) = x^t_r$
    \item The client computes  $k_1= F.Puncture(sk_i,x^t)$
    \item If the hint is marked with $+y$ and possibly other negative constraints
    \begin{itemize}
        \item The client computes \ $k_2 = F.Resample(sk_i,k_1,+y,-c^1,\dots,-c^p)$.
        \item The client sends $k_2$ to the server.
    
    \end{itemize}
    \item Else, the client sends $k_1$ to the server.
    \end{itemize}
    
     \item \textbf{Refreshing}: Same as in \textit{Hyb4} :  
     \begin{itemize}
     \item The client samples a new key $sk'$, and replace the used hint with a new one : $(sk',{+x^t})$
        \item The client then marks the constraint \( -x^t \) for the \( i-1 \)-th previous hints.
     \end{itemize}
    
\end{itemize}
\paragraph{Proof.}
The proof is similar to PPPS \cite{Efficient}. \\
As previously done, by defining the matched hint vector, we can define an equivalent experiment for \textit{Real}, where the client receives a query vector $(x^1,\dots, x^t)$ and, based on it, generates hints conditioned on this vector. Then, the matched hint vector $(i_1,\dots,i_t)$ is generated, and the corresponding set is returned.

Similarly, we define an equivalent experiment for \textit{Hyb4}, where a matched hint vector is sampled following the same distribution as in the previous experiment. Then, the hints are generated based on the matched hint vector.

A key observation is that in \textit{Hyb4}, the adversary cannot observe the hint table before round $t$ (because it is generated from the matched hint vector). From the adversary’s perspective, the entire table’s distribution is the posterior distribution after observing $i_1, \dots, i_t$. Furthermore, this posterior distribution is exactly the same as the one recorded by the client as constraints. Thus, conditioned on the same matched indices $i_1, \dots, i_t$, the selected entries in both experiments follow the same distribution. Consequently, the adversary’s view in both experiments remains computationally indistinguishable.

The \textit{Real*} experiment, from the server's perspective, behaves exactly like the real protocol, except that we have removed the part related to correctness.  
Therefore, we can conclude that the \textit{Ideal} experiment and the \textit{Real} protocol are computationally indistinguishable, thus completing the privacy proof.

\subsection{Correctness Proof}
We assume that $n$ is bounded by $\text{poly}(\lambda)$ and $\text{poly}(\kappa)$, and let $\alpha(\kappa)$ be a super-constant function, i.e., $\alpha(\kappa) = \omega(1)$. Setting:

\[ M_1 = \sqrt{n}\log\kappa \cdot \alpha(\kappa), \quad M_2 = \log{\kappa} \cdot \alpha(\kappa). \]

All queries $Q = \sqrt{n}/2$ will be answered correctly with probability at least $1 - \text{negl}(\lambda) - \text{negl}(\kappa)$ for some negligible function $\text{negl}(\cdot)$.

We assume that the $Q$ queries are distinct and random, meaning they are sampled randomly from $\{0,1,\dots,n-1\}$ without replacement. 

There are two types of failure events:
\begin{itemize}
    \item The client cannot find a set that contains the queried index in the primary table.
    \item The client runs out of hints in the backup group.
\end{itemize}
Let's look at the first type of event: 
For any query $x^t$, the probability of the event "the query fails because no hint in the primary table contains $x^t$" (A) can be written as:

\[ P(A) = \prod P(G_{x^t_l}(sk_i) \neq x_r), \]

by independence of the key samples.
This independence holds because, at any given time, each hint in the primary table (even when some hints have been replaced by backup ones), generates a set of indices that appears uniformly random (even when we take into account the constraints) and pairwise independent, as proven in the privacy proof.
Thus:

\[ P(A) = \left(1 - \frac{1}{\sqrt{n}}\right)^{M_1}. \]

And then: 

\[ \left(1 - \frac{1}{\sqrt{n}}\right)^{M_1} \leq e^{-\frac{M_1}{\sqrt{n}}} = e^{-\log{\kappa} \alpha(\kappa)} = \kappa^{-\alpha(\kappa)}. \]

Since $\kappa^{-\alpha(\kappa)}$ is negligible in $\kappa$, this proves correctness for the first type of failure. \\
Let's now look at the second type of event : \\
For a chunk $k$, we need to prove that the client does not easily run out of queries. This only happens when the client makes more than $M_2$ queries in a single chunk.

Since the client makes $\sqrt{n}/2$ queries, and there are $\sqrt{n}$ groups, we define the random variables $Y_{t,i} \in \{0,1\}$ such that $Y_{t,i} = 1$ if and only if the $t$-th query is located in the $i$-th chunk. We define:

\[ X_i = Y_{1,i} + \dots + Y_{Q,i}, \]

which represents the number of queries in chunk $i$. We know:

\[ E[Y_{t,i}] = \frac{1}{\sqrt{n}}, \quad E[X_i] = 1/2. \]

Using the Chernoff bound, we have 
\[P(X_i> (\log(\kappa)\alpha(\kappa)) \le \kappa^{-\alpha(\kappa)}E(e^{X_i})
\]
Notice that we rely on the randomness of the permutation and the queries do not have duplication, so $Y_{i,1},\dots,Y_{i,Q}$ are negatively correlated, therefore : 
\[E[e^{X_i}] \le \prod_jE(e^{Y_{i,j}})  = \prod_j (1+\frac{e-1}{\sqrt{n}})
= (1 +\frac{e-1}{\sqrt{n}})^{\sqrt{n}/2} \le e^{\frac{e-1}{2}}\]

And finally : 
\[
 P(X_i> (\log(\kappa) \cdot\alpha(\kappa)) \le e^\frac{e-1}{2}\cdot\kappa^{-\alpha(\kappa)}
\]

Taking the union bound over all $\sqrt{n}$ chunks, the failure probability is bounded by:

\[ \sqrt{n} \cdot e^\frac{e-1}{2}\kappa^{-\alpha(\kappa)}, \]

which is a negligible function of $\kappa$ because $n$ is bounded by $poly(\kappa)$. This completes the proof of correctness.

\subsection{Efficiency} \label{subsec:efficiency}
\paragraph{Client computation}
Let's suppose first that the client finds a hint with no positive constraint : 
Then the protocol followed is exactly the one described in treePIR, which has been proved to run in $O(\log{n}\sqrt{n})$ time. \\
Now, we consider the case where the client has to enforce some constraints.  The only change is that the client must run the Resample algorithm. We define "a check" when you have to check the value of a leaf of the tree which takes exactly $\log{n}/2$ calls to the PRG $G$.
Let's denote the probability that every negative constraint is satisfied $P_{neg}$. Because we have at most $\sqrt{n}/2$ queries, we have at most $\sqrt{n} /2$ constraints denoted $c^1, \dots, c^{\sqrt{n}/2}$ Then:  
\[
P_{neg} = 1 - P\left(\bigcup c^i \text{ is not respected} \right)
\]
\[
P_{neg} > 1 - \sum P(c^i \text{ is not respected}) = 1 - \frac{\sqrt{n}}{2\sqrt{n}} = \frac{1}{2}
\]

The probability is at least $1/2$ so $O(\sqrt{n})$ checks are needed for negative constraints.\\
Also, since the probability that the positive constraint is enforced is $\frac{1}{\sqrt{n}}$, the positive constraint requires \( O(\sqrt{n}) \) checks. Since we only verify the negative constraints after the positive constraint, the resampling process performs \( O(\sqrt{n}) \) checks. Each check takes \( O(\log{n}) \) PRG-calls to execute. Consequently, the full resampling process runs in \( O(\log{n} \sqrt{n}) \) PRG-calls. This implies that resampling does not introduce any additional asymptotic complexity, compared to the case where resampling is not involved. Therefore, the client runs in $O(\log{n}\sqrt{n})$ PRG-calls for every query.\\
There is one final computation to consider: the client must construct the tables required for the next \( \sqrt{n}/2 \) queries. To achieve this, upon receiving a chunk \( c \) along with the query response, the client must:  

\begin{itemize}
    \item For each entry in the next primary table \( (sk, p_{sk}, ()) \), update \( p_{sk} \) as follows:  
    \[
    p_{sk} \leftarrow p_{sk} \oplus DB[P.eval(sk, c)].
    \]
    This requires \( O(\log{\sqrt{n}}) \) PRG calls and a single XOR operation.  

    \item For each entry in the next backup table \( (sk, p_{sk}, (\gamma_1, \dots, \gamma_{\log(\sqrt{n})})) \), update both \( p_{sk} \) and, if necessary, the values of \( \gamma \). This step involves \( O(\log{n}) \) PRG calls and at most \( O(\log(n)) \) XOR operations.  
\end{itemize}  

Taking into account the computation required for constructing the new primary and backup tables, the overall client computation becomes  $O_\lambda(\sqrt{n} \log{n} \log{\kappa} \cdot \alpha(\kappa))$.

\paragraph{Server computation}
The proof in TreePIR shows that the server computation is in $O_\lambda(\sqrt{n}(\log{n})^2)$ time. 
In our protocol, the server makes the same computation but in addition it also has to send back intermediate calculations (the parity values of the subkeys). However, those parities were already computed in the TreePIR server protocol, so it does not introduce any additional computation complexity. 
Therefore, the server runs in $O_\lambda(\sqrt{n}(\log{n})^2)$.
\paragraph{Client Storage}
We prove that the storage needed for the client is \( O_\lambda(\sqrt{n} \log{n} \log{\kappa} \cdot \alpha(\kappa)) \).  

For the primary table, each entry consists of a secret key, an xor-value, and a set of constraints. However, we do not store the constraints explicitly. Instead, the positive constraint is embedded in the hint, while the negative constraints are managed using an auxiliary array. This array stores tuples of the form \( (y, i) \), where \( y \) represents the constraint that must be satisfied, and \( i \) is the index of the hint used. For instance, when resampling is required for a hint of index $j$ we make sure to enforce constraints that come from a hint whose index was higher than $j$. Since there are at most \( O(\sqrt{n}) \) constraints, this does not introduce additional storage.

However, for the backup tables, we must also account for the parities of the subtrees discussed earlier. This adds \( O(\log n) \) additional parity values to consider per hint. Given that there are \( O(\sqrt{n} \log{\kappa}\cdot \alpha(\kappa)) \) hints of this type, the total client storage requirement ultimately amounts to \( O_\lambda(\sqrt{n} \log{n} \log{\kappa}\cdot\alpha(\kappa)) \).

\subsection{Performance Summary}
The proposed scheme achieves:
\begin{itemize}
    \item $\mathcal{O}_\lambda(\sqrt{n}\log{n}\log{\kappa } \cdot \alpha(\kappa))$ client storage and no additional server storage.
    \item \textbf{Preprocessing phase}
    \begin{itemize}
        \item $\mathcal{O}_\lambda(n \log{n} \log{\kappa} \cdot\alpha(\kappa))$ client time and $\mathcal{O}(n)$ server time. 
        \item $\mathcal{O}(n)$ communication
    \end{itemize}
    \item \textbf{Query phase}
    \begin{itemize}
        \item $\mathcal{O}_{\lambda}(\sqrt{n}\log{n}\log{\kappa} \cdot \alpha(\kappa))$ client computation
        \item $\mathcal{O}_\lambda((\log{n})^2\sqrt{n})$ server computation
        \item Bandwidth: $\mathcal{O}_\lambda(\log{n})$ upload and $\mathcal{O}(\log({n}) \sqrt{n})$ download
    \end{itemize}
\end{itemize}
\section{Bibliography}
\bibliographystyle{plain}
\bibliography{reference}
\end{document}